\documentclass[
 prc,
 preprint,
 showpacs,
 amsmath,amssymb,
 aps,floatfix,
 showkeys,
 superscriptaddress]{revtex4}
\usepackage{graphicx}
\usepackage{bm}
\begin{document}
\title{
Reaction cross sections for proton scattering from stable and 
unstable nuclei based on a microscopic approach}
\author{H. F. Arellano}
\email{arellano@dfi.uchile.cl}
\homepage{http://www.omp-online.cl}
\affiliation{Department of Physics - FCFM, University of Chile\\
             Av. Blanco Encalada 2008, Santiago, Chile}
\affiliation{
Commisariat \`a l'Energie Atomique,
D\'epartement de Physique Th\'eorique et Appliqu\'ee,
Service de Physique Nucl\'eaire,
Boite Postale 12, F-91680 Bruy\`eres-le-Ch\^atel, France}
\author{M. Girod}
\affiliation{
Commisariat \`a l'Energie Atomique,
D\'epartement de Physique Th\'eorique et Appliqu\'ee,
Service de Physique Nucl\'eaire,
Boite Postale 12, F-91680 Bruy\`eres-le-Ch\^atel, France}

\begin{abstract}
Microscopic optical model potential results for reaction cross
sections of proton elastic scattering are presented. 
The applications cover the 10-1000 MeV energy range and
consider both stable and unstable nuclei.
The study is based on \emph{in-medium} $g$-matrix full-folding
optical model approach with the appropriate relativistic kinematic 
corrections needed for the higher energy applications.
The effective interactions are based on realistic \emph{NN} potentials
supplemented with a separable non-Hermitian term to allow
optimum agreement with current \emph{NN} phase-shift analyzes,
particularly the inelasticities above pion production threshold.
The target ground-state densities are
obtained from Hartree-Fock-Bogoliubov calculations based on the
finite range, density dependent Gogny force.
The evaluated reaction cross sections for proton scattering
are compared with measurements and their systematics is analyzed.
A simple function of the total cross sections in terms of 
the atomic mass number is observed at high energies.
At low energies, however, discrepancies with the available data 
are observed, being more pronounced in the lighter systems.
\\
\end{abstract}



\maketitle

\section{Introduction}
Nucleon-nucleus integrated cross sections constitute a key observable in  
fundamental nuclear research as well as in applications of nucleon-induced 
reactions.
These quantities are of particular importance in nuclear technology 
such as nuclear transmutation, nuclear waste treatment, safety assessment 
and medical therapy, among many examples.
From a fundamental point of view, their description within global
microscopic approaches constitutes a stringent test to the understanding 
of the underlying physics involved in the interaction of nucleons
colliding with a nucleus.
Depending on the energy range of the projectile, total cross sections
may constitute an important input in the study of diverse phenomena.  
For instance, the study of $r$-processes in astrophysics require 
the knowledge of total cross sections at energies near and below a few MeV,
whereas spallation applications may well require data up to GeV energies.

In this article we present a study of the systematics exhibited
by proton-nucleus (\emph{pA}) reaction cross sections 
based on the microscopic full-folding (FF) optical model potential 
approach \cite{Are95,Are02}.
The study spans in energy from a few MeV up to 1 GeV, considering 
various even-even isotopes from carbon up to lead.
The calculated reaction cross sections are compared with existing
data, whereas results for proton scattering from unstable nuclei 
are examined as functions of the number of constituents of the target.

Research on proton reaction cross section of \emph{pA} elastic scattering 
has received significant attention during recent years. 
These studies include experimental \cite{Auc05}, 
phenomenological \cite{Koh05,Ing05}
and microscopic \cite{Amo06,Deb05,Deb03,Deb01} analyses.
Global analyses in the framework of Glauber theory have also been
reported \cite{Vis00}.
Recent experimental efforts by Auce and collaborators \cite{Auc05} 
have been very valuable in reporting new measurements at intermediate 
proton energies. 
The phenomenological studies of Refs. \cite{Koh05,Ing05}
represent attempts to provide simple parametrizations of the
observed cross sections in terms of geometry and mass distribution 
of the targets.
The microscopic studies reported in Refs. \cite{Amo06,Deb05,Deb03} 
are based on the referred `coordinate-space $g$-folding model' \cite{Amo00},
where a \emph{local} medium-dependent \emph{NN} effective interaction is folded
with the target ground-state mixed density.
As a result, a nonlocal optical potential is obtained, where the non locality
stems from the inclusion of the exchange term.
The results reported here differ from the above $g$-folding 
model in two aspects. The first of them is that the antisymmetrized 
\emph{NN} effective 
interaction is handled in momentum space with no special considerations on 
an eventual local structure in coordinate space\cite{Are95}. 
Thus, we retain the intrinsic off-shell behavior of the interaction.
The second difference lies in the representation on the mixed density.
Whereas the $g$-folding approach treats explicitly the mixed density
extracted from shell-models,
the results we report here rely on its Slater representation using only
the diagonal elements. 
An assessment of this approximation has been made within the 
free $t$ matrix full-folding approach, where it is 
shown that it affects only slightly the differential scattering 
observables at momentum transfers ($q$) of 1 fm$^{-1}$, becoming 
more visible at $q\geq$ 2.5 fm$^{-1}$ (c.f. Fig. 3 of Ref. \cite{Are90b}).

This article is organized as follows:
In Section II we outline the general framework upon which we base our study.
In Section III we present results for the calculated reaction cross
sections for \emph{pA} elastic scattering at energies between 10 MeV
and 1 GeV.
Additionally, we examine the systematics exhibited by the calculated
reaction cross sections and the total cross sections for
neutron-nucleus (\emph{nA}) elastic scattering at energies above 400 MeV.
Finally, in Section IV we present a summary and the 
main conclusions of this work.

\section{Framework}
The microscopic approach we follow is the
\emph{in-medium} FF optical model approach \cite{Are95},
which is a realization of the double convolution of an \emph{NN} 
antisymmetrized effective interaction with the target 
ground-state mixed density.
In its formulation, with the use of the Slater 
representation of the mixed density and the assumption 
of a weak dependence of the $g$ matrix on one of the momentum 
integrals, the FF optical potential for \emph{pA} collisions 
at beam energy $E$ takes the simplified form 
\begin{eqnarray}
\label{upp}
U_{pp}({\bm k}',{\bm k};E)&=&\int d{\bm Z}\;
e^{i({\bm k}'-{\bm k})\cdot{\bm Z}} 
\sum_{\alpha=p,n}\;\times \nonumber \\
& &
\rho_\alpha({\bm Z})\;\bar g_{p\alpha}[{\bm k}',{\bm k};\rho_{A}(Z)]\;,
\end{eqnarray}
where $\rho_\alpha$ is the point density of specie $\alpha$,
and $\bar g_{p\alpha}$ represents an off-shell Fermi-averaged 
amplitude in the appropriate $p\alpha$ pair and evaluated
at the local target density $\rho_{A}({\bm Z})$.
More explicitly, in an infinite nuclear matter model these
density-dependent amplitudes are given by
\begin{equation}
\label{g_on}
\bar g_{NN}({\bm k}',{\bm k};\bar\rho) =
\frac{3}{4\pi \hat k^3} \int
\Theta(\hat k-|{\bm P}| )
g({\bm k_r}',{\bm k_r};\:\sqrt{s};\:\bar\rho\:)
\;d{\bm P}\;,
\end{equation}
where $g({\bm k_r}',{\bm k_r};\sqrt{s};\bar\rho)$
corresponds to off-shell $g$ matrix elements for 
symmetric nuclear matter of density $\bar\rho$. 
The relative momenta, ${\bm k_r}'$ and ${\bm k_r}$, depend
on the asymptotic momenta ${\bm k}',{\bm k}$ as well as 
${\bm P}$, the mean momentum of the struck target nucleon.
Their general form is ${\bm k_r}= W {\bf k} - (1-W){\bf p}$, with
${\bf p}={\bf P}/2+{\bf k}-{\bf k'}$, and $W=W(E;{\bf k'},{\bf k},{\bf P})$.
As there is no local prescription to handle the $g$ matrix,
the results we report here retain the intrinsic 
non localities of the \emph{NN} interaction.
The $s$-invariant specifies the energy at which the interaction is
evaluated, which depends on the beam energy $E$ and kinematics
of the interacting pair.
Details about the implementation of the Fermi-averaged
elements and relativistic kinematics corrections have been discussed
in Ref. \cite{Are02}.
With the above considerations we obtain a nonlocal optical 
potential which is handled exactly within numerical accuracy.

An important element for the realization of the optical potential 
is the two-body effective interaction, which in our model takes the
form of the nuclear-matter $g$ matrix. 
This is based on the Brueckner-Bethe-Goldstone model for symmetric 
nuclear matter and described by the integral equation
\begin{equation}
\label{bbg}
g(\omega)=V+V\frac{\hat Q}{\omega-\hat e_1-\hat e_2}\;g(\omega)\;,
\end{equation}
with $V$ the free-space two-nucleon potential, 
$\hat e_1$ and $\hat e_2$ the quasiparticle energies,
$\hat Q$ the Pauli blocking operator and $\omega$ the starting
energy just above the real axis.
In the case of the free $t$ matrix applications, the Pauli blocking
operator is set to unity and the nuclear self-consistent fields
vanish. Thus, $g\to t$, the \emph{NN} scattering matrix in free space.

A physically acceptable $g$ matrix for GeV nucleon energy applications 
requires a bare \emph{NN} potential model consistent with its high energy
phenomenology, particularly the presence of complex phase-shifts 
above pion-production threshold \cite{Arn00}.
In our approach this is achieved by supplementing realistic \emph{NN} 
potential models ($V_R$) with a separable term of the form \cite{Are02}
\begin{equation}
\label{sep}
V = V_{R}+\mid\xi\rangle\Gamma(E)\langle\xi\mid\;.
\end{equation}
In this study the form factors $\mid\xi\rangle$ are taken as 
harmonic oscillator states characterized with 
$\hbar\omega=$450 MeV. 
The strength $\Gamma(E)$ is energy and state dependent, becoming 
complex in those states where loss of flux is observed. 
These coefficients are calculated analytically for each 
state to reproduce the year-2000 \emph{np} continuous-energy solution (SP00)
of the phase-shift analysis by R. Arndt \emph{et al.} \cite{Arn00}.
With these considerations we are able to account for the absorption
stemming from the elementary \emph{NN} interactions.
Since most realistic bare potentials do not share a common phase-shift 
data basis, in particular the SP00 solution, we suppress the separable 
contribution below 300 MeV. The inclusion of the separable strength
is done gradually out to 400 MeV, energy from which the full strength is
taken into account.

Another important input in our calculations is the target 
ground-state radial density.  
These are obtained following self-consistent Hartree-Fock-Bogoliubov
calculations with the Gogny force \cite{Dec80,Ber91}. 
This interaction contains a central finite-range term, 
a zero-range spin-orbit contribution and
a zero-range density-dependent contribution.
This approach has been instrumental for obtaining the spherical radial
densities for $^{12}$C and the isotope families  
$^{12-26}$O, $^{34-64}$Ca, $^{50-86}$Ni, $^{80-100}$Zr, $^{96-136}$Sn 
and $^{176-224}$Pb included in this study.

\section{Applications}
\subsection{Scattering from stable nuclei}
\begin{figure}
 \includegraphics[scale=0.70]{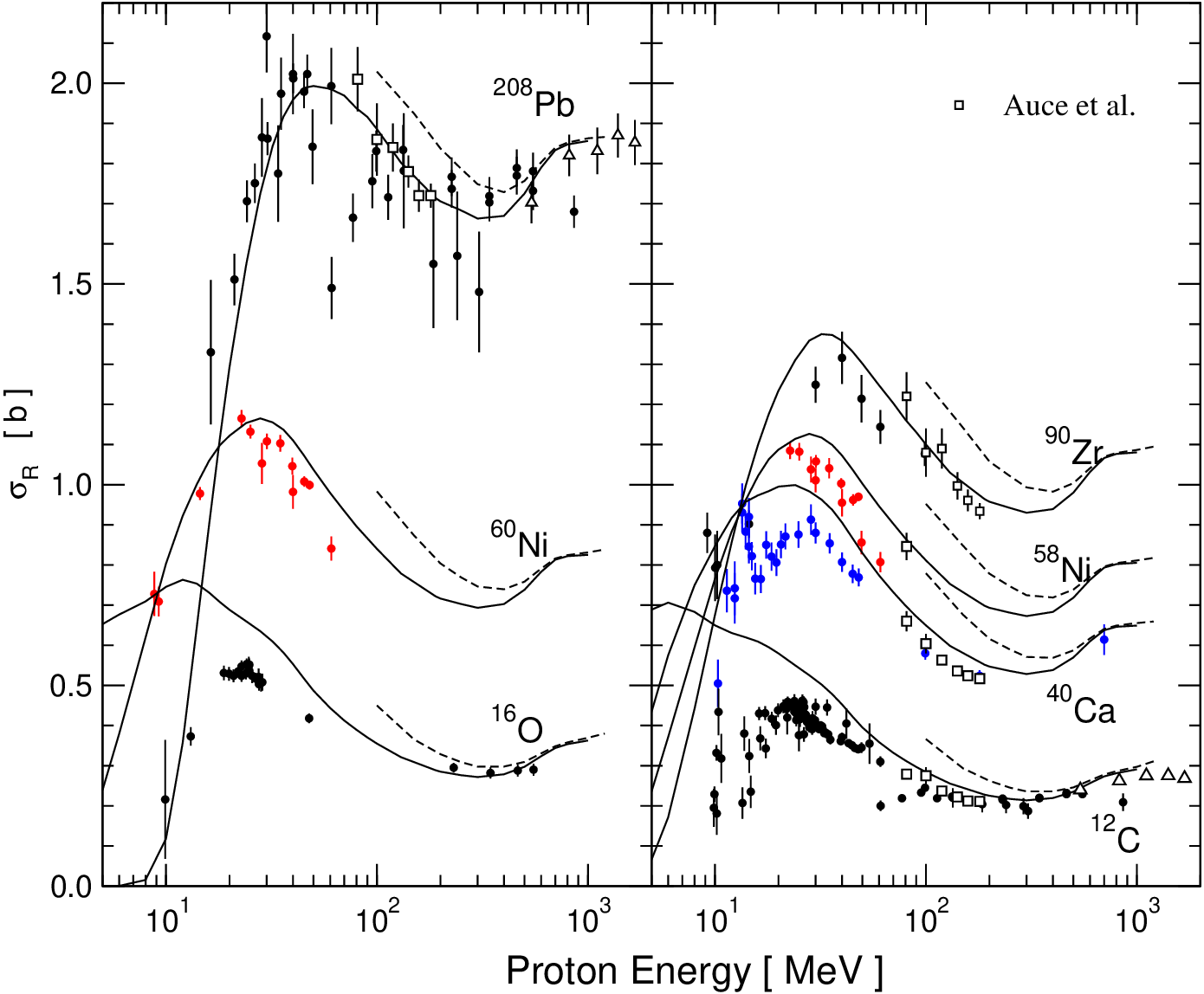} 
\bigskip
\caption{{\protect\small
\label{fig1}
(Color online) Measured \cite{Car96,Auc05,Die02} and calculated 
reaction cross section based on full-folding optical model potential 
using the $g$ matrix (solid curves) and the free $t$ matrix (dashed curves).
}}
\end{figure}

Following the considerations outlined above, we have calculated the 
reaction cross sections for proton-nucleus elastic scattering $\sigma_R$
at beam energies ranging from 5 MeV up to 1 GeV. 
The targets considered are 
$^{208}$Pb, $^{90}$Zr, $^{60}$Ni, $^{58}$Ni, $^{40}$Ca, $^{16}$O and $^{12}$C,
whose proton and neutron root-mean-square (r.m.s.) radii are summarized
in Table \ref{rms}.
The results for $\sigma_R$ are shown in semi-log scale in Fig. \ref{fig1}, 
where the data are taken from Refs. \cite{Auc05} (open squares),
\cite{Car96} (filled circles) and \cite{Die02} (triangles).
The calcium and nickel data have been colored blue and red, respectively,
to distinguish them when they overlap.
The solid curves represent results from the FF optical model using 
the nuclear matter $g$ matrix based on the \emph{np} Argonne V18 reference 
potential \cite{Wir95}, whereas the dashed curves represent results 
using the free $t$ matrix corresponding to the same bare interaction.
As observed, the level of agreement between the calculated cross 
sections and the data is qualitatively different above 200 MeV 
from that below 100 MeV.
At proton energies below 100 MeV --with the exception of $^{208}$Pb-- 
all FF results overestimate $\sigma_R$,
being more pronounced for the $^{12}$C and $^{16}$O targets. 
In these two cases differences of $\sim$100 mb at 30 MeV are observed,
with no proper account of the maxima near 25 MeV exhibited by the data. 
Instead, the calculated cross sections 
decrease monotonically in the range 10--300 MeV.
This feature is also observed in Ref. \cite{Deb01},
where a much closer agreement with the data is reported.
The deficiencies of microscopic optical models at low energies have 
been addressed recently in Ref. \cite{Dup06}, where it is suggested 
that the inclusion of nucleon-phonon couplings may be needed to improve 
the agreement with the low energy data. 

For the heavier targets ($^{40}$Ca, $^{58,60}$Ni, $^{90}$Zr and $^{208}$Pb) 
the calculated cross sections follow the same qualitative trend of the data. 
The discrepancies observed at energies between 20 and 100 MeV 
can be characterized by a uniform 50-mb overestimate of the
calculated cross sections relative to the data.
At energies above 200 MeV the agreement with the measured cross sections
is considerably improved, with the exception of the 860-MeV measurements
of $^{12}$C and $^{208}$Pb, where a clear disagreement is exposed.
At this specific energy we have performed FF optical model calculations
for the targets considered and our predictions are summarized in 
Table \ref{predictions}. 
These predictions are consistent with the data from Ref. \cite{Die02} near
and above 1 GeV (open triangles), an indication of the adequacy of our
approach at these high energies.

\begin{figure}
 \includegraphics[scale=0.70]{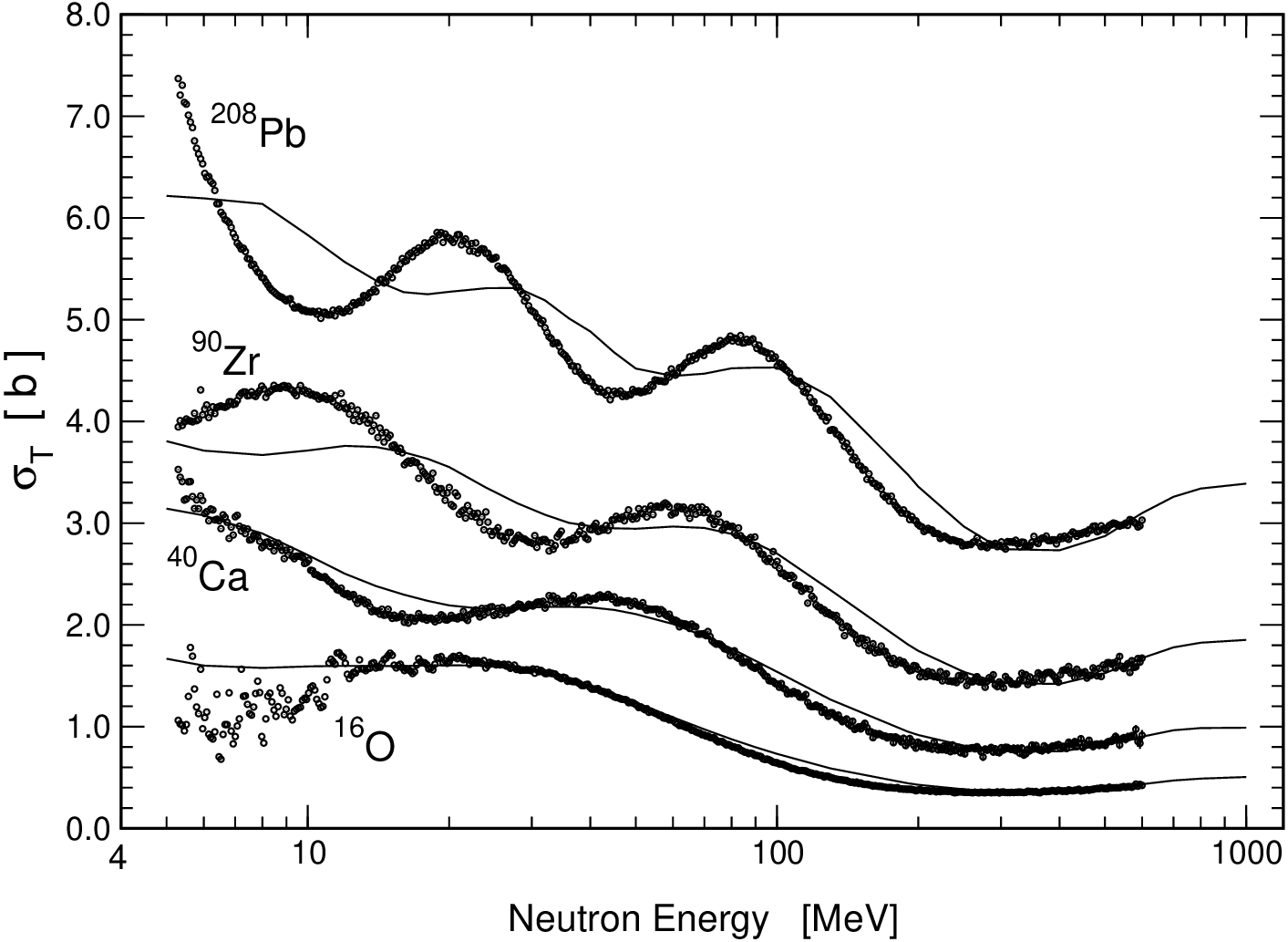} 
\bigskip
\caption{{\protect\small
\label{total}
The measured \cite{Fin93} and calculated total cross section 
based on the \emph{in-medium} FF optical model potential.
}}
\end{figure}

For completeness in this section, in Fig. \ref{total} we compare
the measured \cite{Fin93} and calculated \emph{nA} 
total cross sections ($\sigma_T$).
The calculated results were obtained within the FF optical model approach,
and the applications cover the energy range between 5 MeV and 1 GeV.
What becomes clear from this figure, in contrast to the description
of $\sigma_R$, is that the closest agreement with the data occurs
with the lighter self-conjugate targets. 
Instead, for the isospin asymmetric $^{90}$Zr and $^{208}$Pb systems, 
the calculated $\sigma_T$ lacks the pronounced oscillating pattern 
exhibited by the data. 
Microscopic calculations within the $g$-folding approach have been 
reported \cite{Deb01} to provide an excellent account 
for the total cross-section data in the 60-200 MeV energy range. 
Its extension to 600 MeV has been achieved by means of a simple 
parametric form \cite{Deb04}.
From the prospective of the microscopic FF approach, 
the account for $\sigma_R$ for isospin asymmetric targets 
remains a pending issue.

\subsection{Scattering from unstable isotopes}
\begin{figure}
 \includegraphics[scale=0.70]{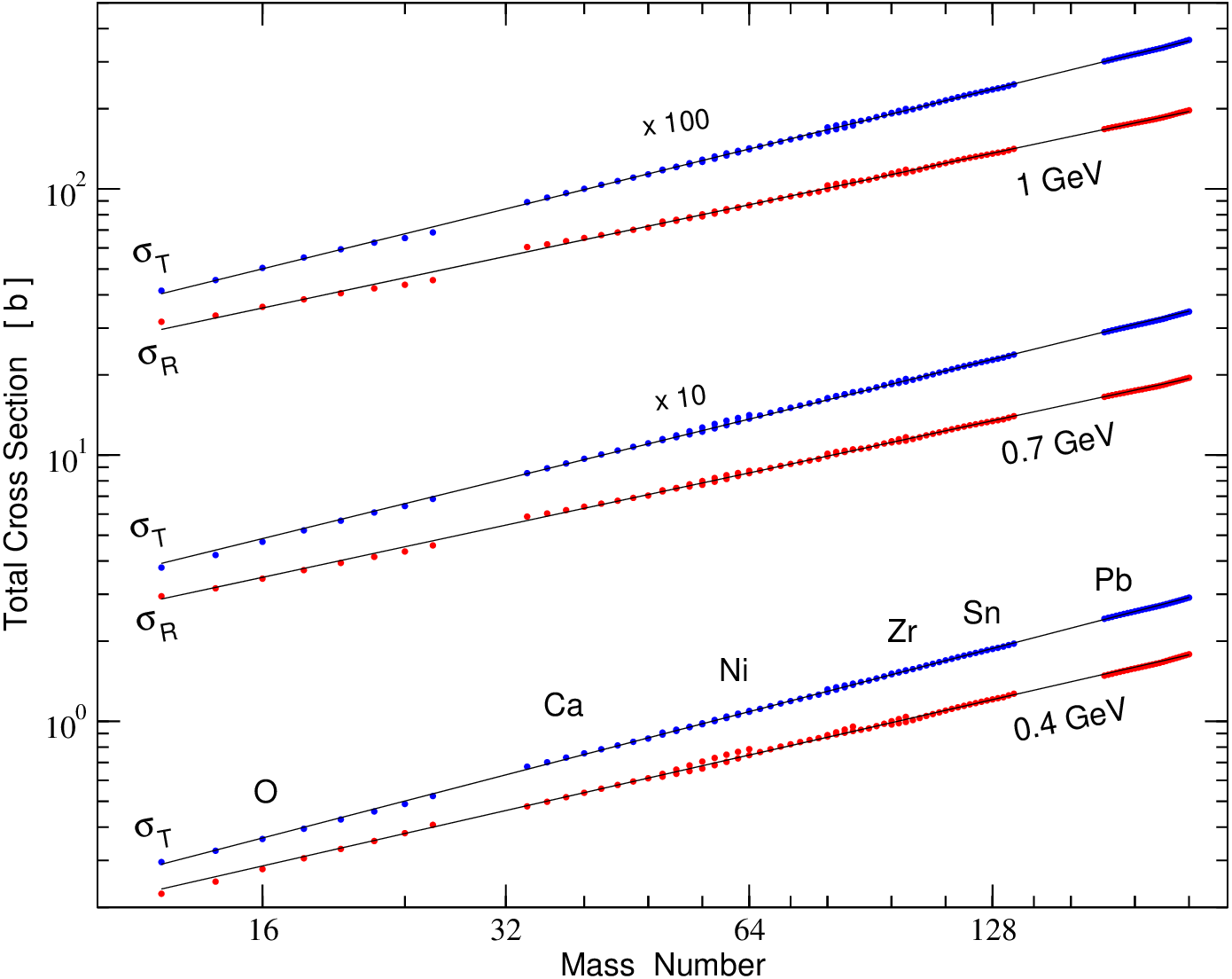} 
\bigskip
\caption{{\protect\small
\label{fig2}
(Color online) Calculated $\sigma_R$ and $\sigma_T$ for nucleon 
elastic scattering from O, Ca, Ni, Zr, Sn and Pb even-even isotopes.  
The lower-, middle- and upper-pair curves correspond to 0.4, 0.7 
and 1.0 GeV nucleon energy, respectively.
}}
\end{figure}

In order to explore the behavior of the integrated cross sections
as functions of the mass number, particularly its isotopic asymmetry,
we have evaluated $\sigma_R$ and $\sigma_T$ for nucleon-nucleus elastic 
scattering.
These applications include results at 0.4, 0.7 and 1.0 GeV nucleon
energies, covering even-even isotopes of oxygen, calcium, nickel, 
zirconium, tin and lead. 
In Fig. \ref{fig2} we show a log-log plot of the calculated $\sigma_R$
(red circles) and $\sigma_T$ (blue circles)
at 0.4, 0.7 and 1.0 GeV nucleon energy as functions of A,
the number of nucleons of the target.
The straight lines represent the least-square regression of the type
\[
\sigma=\sigma_0\;A^{\;p} \;,
\]
with $\sigma_0$ the reduced cross section and $p$ the power-law exponent. 
Both parameters depend on the nucleon energy $E$.
For clarity in the figure, the results for 0.7 GeV and 1.0 GeV have 
been off-set by factors of 10 and 100, respectively.

As observed from the figure, with the exception of $\sigma_R$ for the 
oxygen isotopes, the overall trend of the calculated cross sections
follows very well the $A^p$ power law. 
In the case of the oxygen isotopes at 1.0 GeV, a slight deviation as 
a function of the asymmetry $A-2Z$ is observed. 
In this particular case the maximum deviation of the calculated
$\sigma_R$ with respect to the $A^p$ law is bound by 3\%.
Our estimate is that this manifestation of the isotopic asymmetry
is genuine, although a more reliable evaluation would require a better
handling of the target mixed density.
Such considerations go beyond the scope of the present work.

In Table \ref{table1} we summarize the results from the 
power law regression of the calculated cross sections. 
In all cases we include the standard deviation of the 
obtained reduced cross section and corresponding exponent.
Judging by the standard deviation, the results at 1 GeV exhibit 
the closest agreement with the $A^p$ behavior, case in which
$\sigma_R\sim A^{0.644}$, and $\sigma_T\sim A^{0.748}$. 
Clearly these results differ from the $A^{2/3}$ geometric law,
suggesting the relevance of the hadron dynamics and 
implicit correlations in the collision phenomena.

\begin{figure}
 \includegraphics[scale=0.70]{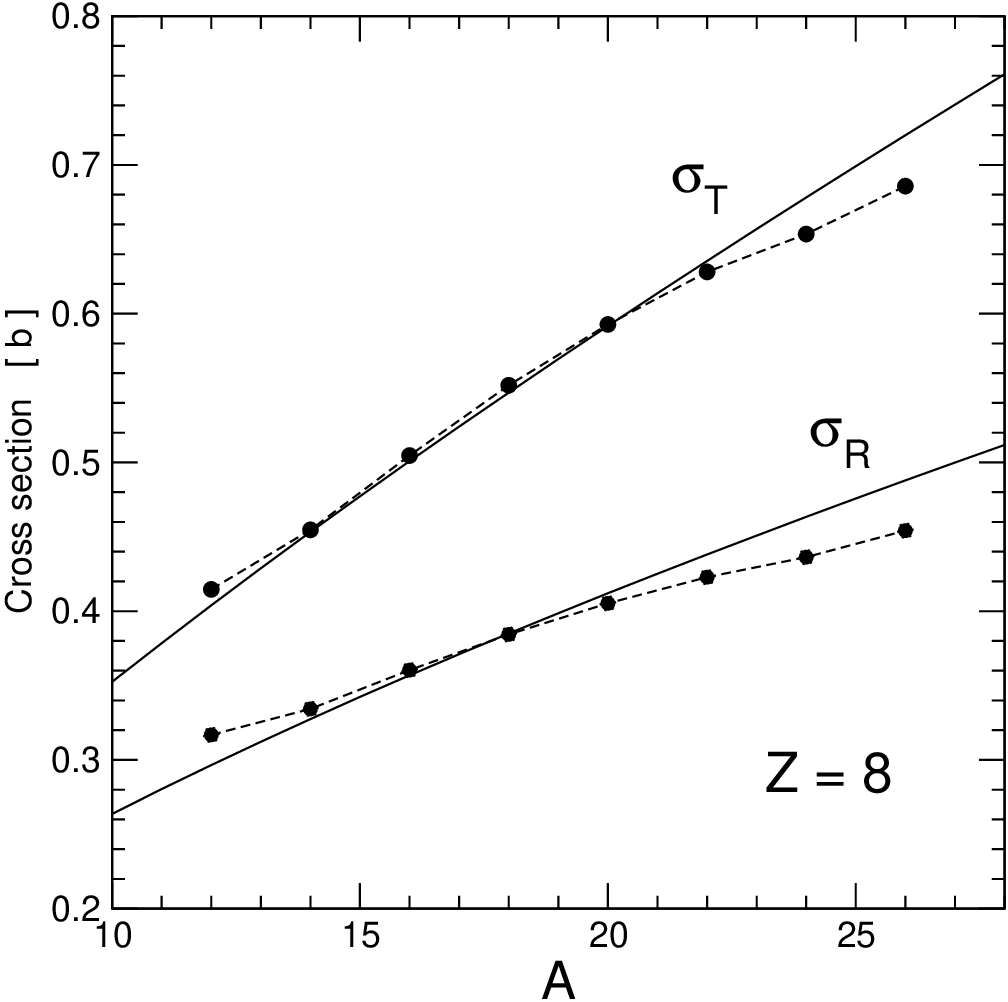} 
\bigskip
\caption{{\protect\small
\label{z8}
The calculated $\sigma_R$ and $\sigma_T$ (filled circles)
\emph{versus} the $A^p$ power law regression (solid curves) for 
1-GeV nucleon scattering from oxygen isotopes.
}}
\end{figure}

A closer comparison between the calculated cross section and 
the least-square power law fit is shown in Fig. \ref{z8},
where the solid curves represent the cross sections for
nucleon scattering from $^{16}$O at 1 GeV using the corresponding 
parameters from Table \ref{table1}.
The filled circles represent the calculated $\sigma_R$ and $\sigma_T$
using the FF approach, where dashed lines are drawn to guide the eye.
In this figure it becomes evident the departure from the $A^p$
behavior when the neutron number varies.
Indeed, the calculated cross sections are weaker than those
obtained from the power law for increasing neutron excess.
Conversely, the oxygen isotopes with fewer neutrons than protons 
yield cross sections greater than the ones prescribed by the 
parametrization.
These features become more pronounced in $\sigma_R$ than in $\sigma_T$.
Although we do not have thorough interpretation of this feature,
we notice that the elementary total cross sections, $\sigma_{pp}$ and
$\sigma_{pn}$, grow very rapidly between 500 MeV and 1.3 GeV nucleon
laboratory energy. 
This increase is more pronounced in $\sigma_{pp}$ than in $\sigma_{pn}$, 
where near 1 GeV $\sigma_{pn}$ is overtaken by $\sigma_{pp}$.
The smaller $\sigma_{pn}$ relative to $\sigma_{pp}$ 
weakens the increase of the total cross section as the number of
neutrons is increased.

\section{Summary and conclusions}
We have presented a global study of the reaction cross section
for proton scattering from various unstable and stable
isotopes at energies between 10 MeV and 1 GeV.
The study is based on the microscopic \emph{in-medium} 
FF optical model potential.
The effective interaction is represented by the nuclear matter
$g$ matrix which is obtained from bare \emph{NN} potentials
with complete account for the loss of flux of the interaction
above pion production threshold.
This feature is implemented with the inclusion of a separable,
energy-dependent component added to the reference potential,
in this case the Argonne AV18 \emph{NN} potential model.
To this purpose, we have used the SP00 phase-shift analysis available
from the George Washington University, Data Analysis Center \cite{Arn00}.
The target ground-state densities were obtained from Hartree-Fock-Bogoliubov 
calculations based on the finite range, density dependent Gogny force.

The calculated $\sigma_R$ for proton scattering from stable nuclei 
are in reasonable description of the data above 200 MeV, but lack 
comparable agreement with the data below 100 MeV. 
These discrepancies become more pronounced in the case of the lightest
targets studied, i.e. $^{12}$C and $^{16}$O, but diminish with increasing size.
At higher energies the FF optical model based on the described
\emph{NN} interactions are consistent with the measured data.
A comparison of the FF optical model approach with the total cross section
data for \emph{nA} scattering shows close agreement in the cases
of self-conjugate targets, but clear disagreement for the isospin asymmetric
nuclei. Thus far we have been unable to identify a microscpic mechanism 
able to account for such discrepancy. 

We have also investigated the systematics of the total cross sections
as a function of the mass number for various isotope families. 
The results for both $\sigma_R$ and $\sigma_T$ for proton and neutron 
scattering, respectively, suggest an $A^p$ power-law accurate within 3\%.
At nucleon energies between 0.7 and 1.0 GeV the exponents for $\sigma_R$
and $\sigma_T$ are close to 2/3 and 3/4, respectively.
Some slight deviation from this law is observed when the number of neutrons
departs from that for the most stable isotope. This feature becomes more
pronounced in the case of $\sigma_R$ at 1 GeV for $^{16}$O, 
but weakens for heavier targets and at lower energies.

The work reported here constitutes a global assessment of the microscopic 
momentum-space FF optical model approach on its account for proton
reaction and neutron total cross section, covering nearly three orders 
of magnitude in energy.
Similar studies have been reported within alternative theoretical 
approaches, such as Glauber theory \cite{Vis00}, the \emph{in-medium} 
$g$-folding approach and global Dirac phenomenology \cite{Deb05}. 
Although some differences occur in the quality of the description of the 
data, particularly with respect to the $g$-folding approach, 
it remains difficult to identify the sources of such differences. 
Indeed, in Ref. \cite{Amo00} the effective interaction is represented
in coordinate space as an expansion of Yukawa form factors of various 
ranges and complex energy dependent strength, 
in the form [c.f. Eqs. (7.1) and (7.2) of Ref. \cite{Amo00}]
\[
t_{eff}^{ST}(r,\omega)=
\sum_{i}\langle\,\theta_i\,\rangle\;t_{eff}^{(i)ST}(r,\omega)
\;,
\]
with
\[
t_{eff}^{(i)ST}(r,\omega)=\sum_{j=1}^{n_i}S_{j}^{(i)}(\omega)
\frac{e^{-\mu_{j}^{(i)r}}}{r}\;.
\]
The calculations reported here do not make any consideration 
regarding the coordinate space structure of the $g$ matrix.
Actually, they are taken directly from the solution of the
Brueckner-Bethe-Goldstone equation.
Additionally, the representation of the mixed density may be
critical in the case of the light systems. 
In this regard, the $g$-folding approach is more detailed
by making explicit use of the full mixed density based on shell
models, a treatment which is pending in the FF approach.

From a broader prospective, the realization of the FF optical model 
relies on a weak dependence on the struck nucleon momentum. 
This assumption is crucial to reduce the number of multidimensional 
integrals by three, making computationally feasible the evaluation of 
optical potentials in the FF and $g$-folding approaches. 
However, the conditions under which this assumption becomes most 
(or least) adequate have not been investigated.
Other aspects such as charge-symmetry of the bare \emph{NN} interaction
and asymmetric nuclear matter effects may also need to be examined.

\begin{acknowledgments}
The authors are indebted to Prof. H. V. von Geramb for providing the 
separable strength needed for the high energy applications of this work.
They also thank J.-P. Delaroche for careful and critical 
reading of the manuscript.
{\small H.F.A.} acknowledges partial funding provided by 
{\small FONDECYT} under grant 1040938.
\end{acknowledgments}


\newpage

\newpage

\newpage

\newpage

\newpage
\begin{table}
\caption{\label{rms}
Root-mean-square radii of the point proton ($R_p$) and neutron ($R_n$) 
densities used in this study. }
\begin{ruledtabular}
\begin{tabular}{c|ccccccc}
Nucleus& $^{208}$Pb&$^{90}$Zr&$^{60}$Ni&$^{58}$Ni&$^{40}$Ca&$^{16}$O&$^{12}$C\\
\hline
 $R_p$ [fm] & 5.437 & 4.210 & 3.717 &3.695 &3.405 & 2.676 &2.419 \\
 $R_n$ [fm] & 5.573 & 4.265 & 3.738 &3.683 & 3.365& 2.656 &2.402 \\
\end{tabular}
\end{ruledtabular}
\end{table}

\newpage
\begin{table}
\caption{\label{table1} 
Results for the $A^p$ power law regression of the calculated 
cross sections as a function of the nucleon energy.}
\begin{ruledtabular}
\begin{tabular}{c|cc|cc}
  & \multicolumn{2}{ c|}{$\sigma_R$}
  & \multicolumn{2}{ c }{$\sigma_T$}\\
 \hline
$E$  [GeV] & $\sigma_0$ [mb] & $p$ & $\sigma_0$ [mb] & $p$\\
\hline
0.4 & 42.0 $\pm$ 0.1& 0.692 $\pm$ 0.003& 40.9 $\pm$ 1.1 & 0.789 $\pm$ 0.005 \\
0.7 & 57.0 $\pm$ 0.8& 0.652 $\pm$ 0.002& 61.4 $\pm$ 0.8 & 0.746 $\pm$ 0.002 \\
1.0 & 59.8 $\pm$ 0.1& 0.644 $\pm$ 0.002& 63.0 $\pm$ 0.1 & 0.747 $\pm$ 0.001 
\end{tabular}
\end{ruledtabular}
\end{table}

\newpage
\begin{table}
\caption{\label{predictions}
Measured \cite{Car96} and predicted $\sigma_R$ for \emph{pA} scattering 
at 860 MeV.}
\begin{ruledtabular}
\begin{tabular}{r|c|c}
\hline
Target & Measured $\sigma_R$ [ mb ] & Predicted $\sigma_R$ [ mb ]\\
\hline
$^{208}$Pb &  1680 $\pm$ 40 & 1852 \\
$^{90}$Zr  &   ---          & 1079 \\
$^{60}$Ni  &   ---          &  823 \\
$^{58}$Ni  &   ---          &  804 \\
$^{40}$Ca  &   ---          &  647 \\
$^{16}$O   &   ---          &  357 \\
$^{12}$C   &   209 $\pm$ 22 &  284 \\
\hline 
\end{tabular}
\end{ruledtabular}
\end{table}

\end{document}